\documentclass{article}

\usepackage{arxiv}

\usepackage{anyfontsize} %
\usepackage{natbib}
\usepackage{colortbl}   
\usepackage{xcolor}     
\usepackage{bm}
\usepackage{steinmetz}

\usepackage[utf8]{inputenc} 
\usepackage[T1]{fontenc}    
\usepackage{url}            
\usepackage{booktabs}       
\usepackage{nicefrac}       
\usepackage{microtype}      
\usepackage{natbib}
\usepackage{doi}
\usepackage{xcolor}

\usepackage{cancel}
\usepackage{graphicx}
\usepackage{enumitem}
\usepackage{tabularx}

\usepackage{comment}

\usepackage{multirow}%
\usepackage{physics}
\usepackage[title]{appendix}%
\usepackage{xcolor}%
\usepackage{textcomp}%
\usepackage{manyfoot}%
\usepackage{booktabs}%
\usepackage{algorithm}%
\usepackage{algpseudocode}%
\usepackage{listings}%

\usepackage{amsmath, amsfonts, amssymb, amsthm, mathrsfs, bm}

\usepackage{hyperref} 
    \hypersetup{
    	colorlinks=true,
    	linkcolor=blue,
    	filecolor=blue,      
    	urlcolor=blue,
    	citecolor=blue
    }

\title{Bayesian Computation of Copula-based Multivariate Models with an Application to Doping Detection}

\date{November 2026}	

\author{Emanuele Masillo\\
	MEMOTEF, Sapienza Università di Roma (IT)\\
        \texttt{emanuele.masillo@uniroma1.it} \\
	\And
	Eva Pigi Vyltanioti\\
	MEMOTEF, Sapienza Università di Roma (IT)\\
    AUEB University of Athens \\
	\texttt{pigieva.vyltanioti@uniroma1.it}\\
	 \AND
	 \href{https://orcid.org/0000-0003-2501-8795}{\includegraphics[scale=0.06]{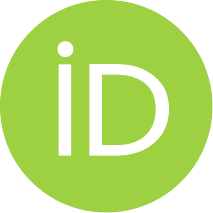}\hspace{1mm} Nina Deliu}  \\
 MEMOTEF, Sapienza Università di Roma (IT)\\
 MRC – Biostatistics Unit, University of Cambridge (UK) \\
	 \texttt{nina.deliu@uniroma1.it} \\
	 \And
     \href{https://orcid.org/0000-0003-2089-3371}{\includegraphics[scale=0.06]{orcid.pdf}\hspace{1mm} Brunero Liseo}  \\
	MEMOTEF, Sapienza Università di Roma (IT)\\
	 \texttt{brunero.liseo@uniroma1.it} \\
}

\renewcommand{\headeright}{ }
\renewcommand{\undertitle}{ }
\renewcommand{\shorttitle}{Bayesian Computation of Copula-based Multivariate Models with an Application to Doping Detection}

\hypersetup{
pdftitle={Bayesian Computation of Copula-based Multivariate Models with an Application to Doping Detection},
pdfsubject={q-bio.NC, q-bio.QM},
pdfauthor={Emanuele Masillo, Pigi Eva Vyltanioti, Nina~Deliu, Brunero~Liseo},
pdfkeywords={Bayesian hierarchical modeling; MCMC Copula estimation; Highest-density regions; Posterior predictive density; Reference ranges.},
}

\begin{document}
\maketitle

\begin{abstract}
The Athlete Biological Passport (ABP) monitors athletes for doping by comparing their biomarker readings against personalized reference ranges. These are estimated via the Bayesian ADAPTIVE model, which analyzes biomarkers separately and ignores their dependence. Building on a recently proposed copula-based multivariate extension of ADAPTIVE, this paper focuses on its computational aspects, introducing a fully joint MCMC algorithm and comparing it with a two-step Bayesian IFM estimator. Simulations show the two procedures perform comparably under a correctly specified copula, but exhibit a genuine trade-off under misspecification: the Bayesian IFM protects marginal inference, while the joint MCMC yields better-calibrated dependence estimates and is markedly faster. 
The two approaches are tested on a dataset containing ABP profiles of elite athletes, provided by NADO Italia, and collected as part of the project {\it Enhancing the methodological framework for doping detection: from univariate to multivariate testing analysis on Athlete Biological Passport profiles}. Both approaches achieve consistent classification performance, providing anti-doping organizations with two concrete inferential routes for adopting a multivariate ADAPTIVE model, along with guidance on the trade-offs between them.
\end{abstract}


\section{Introduction}\label{sec: Intro}

Copula models are flexible tools for representing complex structures of dependence for random vectors. According to Sklar's theorem \citep{sklar_fonctions_1959}, any multidimensional absolutely continuous distribution function can be uniquely represented as a copula, i.e. a joint cumulative distribution function on the unit hypercube with uniform marginals. The copula representation captures the dependence structure among the vector components, disentangling it from the marginal behaviour. 

From a statistical perspective, interest is typically focused on estimating the copula parameters, a task complicated by the dependence on the estimation of the marginal parameters. In the classical (frequentist) framework, this issue is commonly addressed using the so-called {\it inference functions for margins} (IFM) estimator~\cite{joe1996estimation}. More in detail, let $\underline{Y}=(Y_1, \cdots, Y_d)$ be an absolutely continuous random vector with CDF $F_{\underline{Y}}(\cdot; \bm{\theta})$; following Sklar's theorem, this can be represented as
\begin{equation}
\label{eq: sklar1}
F_{\underline{Y}}(\underline{y}; \bm{\theta}) = 
C\left (F_1(y^{(1)}; \lambda_1), F_2(y^{(2)}; \lambda_2), \dots ,
F_d(y^{(d)}; \lambda_d); \bm{\psi}\right),
\end{equation}
where $C$ is a CDF, the parameter vector $\bm{\theta} = (\bm{\lambda}, \bm{\psi})$
is partitioned into the components associated with the marginals
$\bm{\lambda}=(\lambda_1, \dots, \lambda_d)$ and those characterizing the copula,
$\bm{\psi}$. The two-step approach of~\cite{joe1996estimation} proposes to estimate each marginal component of $\bm{\lambda}$ independently via maximum likelihood estimation (MLE) using only the marginal distributions of the data, and then the dependence parameter $\bm{\psi}$ through a
profile-likelihood-type approach. This is obtained by re-expressing Eq.~\eqref{eq: sklar1} with $\bm{\lambda}$ replaced by its conditional MLE, say $\hat{\bm{\lambda}}_{\bm{\psi}}$. While this approach yields consistent estimates under general conditions, it is known to generate overly confident interval estimates by neglecting the standard error of $\hat{\bm{\lambda}}_{\bm{\psi}}$.

Recently, \citep{deliu2025multivariatecopulabasedbayesianframework} have proposed a Bayesian version of the IFM approach, similarly modulated through two steps. In the first step, the marginal parameters $\bm \lambda$ are estimated in a Bayesian fashion, leading to a chain of posterior draws. In the second step, for each posterior draw of the marginal parameters, posterior samples are also obtained for the copula parameter $\bm \psi$, fully propagating in this way uncertainty from the marginal to the copula estimation. 
It is interesting to note how this approach can be considered a special case of the broader class of \emph{modular}, or \emph{cut}, Bayesian inference schemes \citep{jacob2017better, pompe2025asymptotics, smith2025cutting}.

The aim of this work is to propose, instead, an alternative to such two-step approaches, following a fully joint Markov chain Monte Carlo algorithm (joint MCMC), which updates marginal and copula parameters jointly within a single chain. By simultaneously exploring the $\bm{\lambda}-$ and $\bm{\psi}-$components of the parameter space, with uncertainty allowed to flow from and to each of the two components, this approach is expected to provide a more reliable measure of uncertainty on the final inference. We are specifically interested in evaluating advantages and disadvantages of adopting such a joint Bayesian approach compared with the Bayesian IFM in \cite{deliu2025multivariatecopulabasedbayesianframework}.

Our methodological contribution is not purely abstract: it is directly motivated by the problem of doping detection in professional athletes. Doping control, or testing, stands at the cornerstone of anti-doping organizations to uphold the integrity of clean sports competitions. Since its establishment in 1999, the World Anti-Doping Agency (WADA) has spearheaded the global campaign to achieve doping-free sports competitions internationally and nationally, with the support of national governments~\citep{houlihan_world_2019, wada_world_2021}. In 2009, the Athlete Biological Passport (ABP) was introduced for monitoring elite athletes in Olympic and Paralympic disciplines that require greater physical effort  ~\citep{sottas_athlete_2011}. 
The main objective of the ABP is to monitor the profiles of individual athletes over time with respect to selected biomarkers that are expected to indirectly unveil the effects of doping. These include markers of the steroid module, which inform about any use of anabolic androgenic steroids, and markers of the hematological module, which focus on identifying potential substances used for the enhancement of oxygen transport or delivery. Within the hematological module of the ABP, WADA~\citep{wada_world_2021} identifies two primary markers to detect atypical profiles. These markers, considered the most specific indicators of blood doping \citep{wada_world_2021}, are hemoglobin concentration (\texttt{Hgb}) and the stimulation index OFF-score (\texttt{OFFs}). The latter is a derived biomarker obtained by combining \texttt{Hgb} with the percentage of reticulocytes in blood (RET\%) according to the following formula: $OFFs = 10 \cdot Hgb - 60\sqrt{RET\%}$. If one of these primary markers takes on a value that falls outside the intra-individual reference range constructed for that athlete at that time, 
the observation is classified as atypical, indicating potential doping violations and prompting further investigation.

The methodological framework underlying ABP is called ADAPTIVE \citep{sottas_bayesian_2007} and it is based on a Bayesian approach that combines population and individual data to define personalized, time-updated reference ranges. However, ADAPTIVE analyzes biomarkers separately, following a univariate approach and without considering their dependence structure. 
A few works have investigated a multivariate extension to the problem, including \cite{eleftheriou_multivariate_2023} and \cite{deliu2025multivariatecopulabasedbayesianframework}. While in \cite{eleftheriou_multivariate_2023} a multivariate Gaussian model is investigated, \cite{deliu2025multivariatecopulabasedbayesianframework} propose a more flexible model, capable of separately representing marginal distributions and dependence structure via copulas \citep{nelsen_introduction_2006}. This approach allows us to capture complex dependencies not adequately described by multivariate Gaussian models. Furthermore, the method in \cite{deliu2025multivariatecopulabasedbayesianframework} extends traditional univariate reference ranges to multivariate highest-predictive regions (HPRs), addressing the challenges of multidimensional data with complex structures. Such predictive reference regions account for previous historical readings of each individual athlete in order to provide tailored tolerances that might flag for potential anomalies. Their construction is based on the work of \cite{hyndman_computing_1996} and recent developments proposed by \cite{deliu_alternative_2024}. This paper builds on the same copula-based model and HPR approach proposed in \cite{deliu2025multivariatecopulabasedbayesianframework}, contributing from a computational perspective with the proposal of a joint posterior sampling scheme and an in-depth comparison to the Bayesian IFM. 

The methodological developments of this work are then applied in our motivating application, using data consisting of original ABP profiles collected by the National Anti-Doping Organisation Italy (NADO Italia). Data consider both in-competition and out-of-competition samples, from national- and international-level athletes competing in various sports disciplines. Only samples from athletes who provided explicit consent to the use of their personal data for this specific research purpose are included. All data have been appropriately anonymized in compliance with privacy disclosure requirements.


The remainder of this manuscript is structured as follows. In Section~\ref{sec: problem}, we briefly formalize the doping-detection problem within a statistical framework and review WADA's state-of-the-art implementation. Focusing on the hematological module and its two primary markers, in Section~\ref{sec: copula_mvt} we describe our multivariate model, where the potential dependence among biomarkers is explicitly accounted for using a copula representation. 
Section~\ref{sec: Bayes_comp} is the core of this paper, where we discuss in detail the Bayesian IFM approach (see Section~\ref{sec: Bayes_IFM}) and then introduce, in Section~\ref{sec: full_MCMC}, the joint MCMC algorithm for producing a posterior sample for the two sets of parameters $\bm{\theta=(\lambda, \psi)}$. The two approaches are empirically compared both in terms of simulations (Section~\ref{sec: simulations}) and an application to the dataset provided by NADO Italia (Section~\ref{sec: application}). These consist of $n = 223$ observations for a cohort of 23 elite athletes, including information on sport and discipline, age, sex, altitude exposure, in- or out-of-competition status, and biomarker measurements. Section~\ref{sec: concl} presents conclusions and future research directions.

 \section{Problem formulation and existing approaches}\label{sec: problem}

In this section, we will briefly describe the problem of identifying anomalous behavior in the analysis of data from ABP registers and formalize it from a statistical point of view.

The methods currently in use and some recent proposals appearing in the literature will be described.

\subsection{State-of-the-art: the univariate ADAPTIVE} \label{sec: uvt}

Consider $n$ athletes indexed by $i$; for each athlete, $t_i$ repeated measurements are available. 
Let $Y^{(\ell)}_{i,j} \in \mathbb{R}$ denote the outcome of the bio-marker $\ell$ measured on subject $i$ on occasion $j$ ($i = 1,\dots, n$;  $j = 1,\dots, t_i$). To accommodate the problem of interest, we assume that the outcome is represented by the two continuous primary markers of the ABP hematological module, with $\ell \in \{\texttt{Hgb}, \texttt{OFFs}\}$. An athlete $i$ may be flagged as suspected if the analysis of a new blood sample of these two markers, say $\mathbf{Y}_{i,t_i+1} = (Y^\texttt{Hgb}_{i,t_i+1}, Y^\texttt{OFFs}_{i,t_i+1})$ reveals an atypical finding relative to the historical readings $\underline{\mathbf{Y}}_{i,t_i} = (\mathbf{Y}_{i,1}, \dots, \mathbf{Y}_{i,t_i})$. 
Often, $\mathbf{Y}_{i,j}$ is accompanied by $p \geq 1$ explanatory variables $\mathbf{X}_{i,j} = [X^{(1)}_{i,j},\dots,X^{(p)}_{i,j}]'$; they represent potentially useful information known to influence $\mathbf{Y}_{i,j}$. In the hematological case, for example, the analysis of elite athletes includes gender, age, ethnic origin, the altitude at which the competition is scheduled, and sport discipline~\citep{sharpe_development_2002}.

Given  the historical path $\underline{\mathbf{Y}}_{i,t_i} \in \mathbb{R}^{2 \times t_i}$ and $\mathbf{X}_{i,j} \in \mathbb{R}^p$, the general problem is to provide a probabilistic framework to estimate the conditional probability distribution, say $p\left(\mathbf{y}_{i,t_i+1}| \underline{\mathbf{y}}_{i,t_i}, \mathbf{x}_{i,t_i+1}\right)$, of a new test $\mathbf{Y}_{i,t_i+1} \in \mathbb{R}^2$. 
This can then be used, combined with a decision rule, to define for each athlete a \textit{tolerance region} of values for $\mathbf{Y}_{i,t_i+1}$: for example, if a future observed reading $\mathbf{y}_{i,t_i+1}$ falls outside a $(1-\alpha)\%$ percentile region, where $\alpha \in (0,1)$ is a predefined \emph{miscoverage} level, then $\mathbf{y}_{i,t_i+1}$ is considered atypical and athlete $i$ may be excluded from the competition. Note that this setting illustrates WADA's paradigm of a \textit{no-start} decision rule based on a future \textit{single} measurement $\mathbf{Y}_{i,t_i+1}$ rather than the \textit{full sequence} $(\mathbf{Y}_{i,1},\dots,\mathbf{Y}_{i,t_i+1})$. The former will be the focus of this work, and we refer to Section 2.2 of~\cite{sottas_forensic_2008} for the complete sequence testing paradigm.

A Bayesian predictive technique has been introduced in \cite{sottas_bayesian_2007} in order to quantify the chance of doping through the use of a hierarchical linear model, where both intra- and inter-individual variability are accounted for.
In that paper, the conditional distribution 
$p\left(\mathbf{y}_{i,t_i+1}| \underline{\mathbf{y}}_{i,t_i}, \mathbf{x}_{i,t_i+1}\right)$ 
is obtained using a parametric Bayesian model where the prior distribution on the unknown model parameters is calibrated based on the information available on that portion of the population most similar to athletes, i.e., "healthy" and young, and the updating step is performed using the continuously observed individual data. In particular, a hierarchical model is considered separately for each primary biomarker $\ell$, assuming that: conditionally on the parameters, (i) an athlete's readings $j\geq1$ are independent over time; (ii) outcomes are independent across athletes $i = 1,\dots,n$. 
Also, for $i= 1,\dots,n$, the following assumptions are made for the within-subject (\textit{intra-individual} variability and the between-subject quantities:
\begin{align}
    Y^{(\ell)}_{i,j} | \mu_i^{(\ell)}, \sigma^{(\ell)} &\overset{ind}{\sim} \mathcal{N}(\mu_i^{(\ell)}, \sigma^{(\ell)}),\quad j\geq1 \notag \\ 
    \mu^{(\ell)}_i | \bar{\mu}^{(\ell)}, \bar{\tau}^{(\ell)} &\sim \mathcal{N}(\bar{\mu}^{(\ell)}, \bar{\tau}^{(\ell)}) \notag \\
    \sigma^{(\ell)} | \bar{\phi}^{(\ell)}, \bar{\psi}^{(\ell)} &\sim \text{Log}\mathcal{N}(\bar{\phi}^{(\ell)}, \bar{\psi}^{(\ell)}),
    \notag
\end{align}
where $\mathcal{N}$ and Log$\mathcal{N}$ denote the Gaussian and the Log-Normal distributions, respectively. 
In the previous expressions, the individual mean $\mu_i^{(\ell)}$ and standard deviation $\sigma^{(\ell)}$ are random and they are assumed to have a common distribution with hyperparameters $\bar{\mu}^{(\ell)}, \bar{\tau}^{(\ell)}$ and $\bar{\phi}^{(\ell)}, \bar{\psi}^{(\ell)}$, respectively. 
These latter quantities are assumed to be known and are selected based on historical information that can depend on the specific characteristics of an athlete $i$. 
After collecting a series of reading $t_i$ for subject $i$, the (univariate) conditional distribution $ p\left(y_{i,t_i+1}^{(\ell)}| \underline{y}_{i,t_i}^{(\ell)}, \mathbf{x}_{i,t_i+1}\right)$ represents the athlete-centred \textit{posterior predictive distribution}:
\begin{equation}
\label{eq: post_pred}
p\left(y_{i,t_i+1}^{(\ell)}| \underline{y}_{i,t_i}^{(\ell)}, \mathbf{x}_{i,t_i+1}\right)
\end{equation}

Notice that while prior knowledge of $(\mu_i^{(\ell)}, \sigma^{(\ell)})$ accounts for \textit{inter-individual} variations only (estimated, e.g., from historical control population data), updates progressively leverage the information from the athlete $y_{i,j}^{(\ell)}, j=1,\dots,t_i$ $i$ only (\textit{intra-individual} variations) to capture his/her unique profile. 


\subsection{Multivariate ADAPTIVE extensions} \label{sec: mvt_Greci}

\subsubsection{A multivariate copula-based approach} \label{sec: copula_mvt}

Restricting attention, as per our application of interest, to a $d = 2$-dimensional outcome vector $\underline{Y} = (Y^{(1)}, Y^{(2)})$, we now consider Sklar's factorization result presented in Eq.~\eqref{eq: sklar1}. In the presence of a vector of explanatory variables $\mathbf{X}$ (e.g., age, sex, altitude), Eq.~\eqref{eq: sklar1} can be extended to a \emph{conditional} copula representation \citep{patton_modelling_2006}. Throughout, we adopt the ``simplifying assumption'' commonly invoked in the vine-copula literature \citep{gijbels_estimation_2015}, whereby the copula linking $Y^{(1)}\mid\mathbf{X}$ and  $Y^{(2)}\mid \mathbf{X}$ does not itself depend on the specific value of $\mathbf{X}=\mathbf{x}$. For a continuous random vector $\underline{Y}$ this implies: 
\begin{align}\label{eq: cond_cop}
    f_{\underline{Y}\mid\mathbf{X}}(y^{(1)}, y^{(2)}\mid \mathbf{x};\bm{\theta}) & =  f_{1\mid\mathbf{X}}(y^{(1)}\mid\mathbf{x}; \bm{\lambda}_1) f_{2\mid\mathbf{X}}(y^{(2)}\mid\mathbf{x}; \bm{\lambda}_2)\\
    & \times c\big(F_{1\mid\mathbf{X}}(y^{(1)}\mid\mathbf{x};\bm{\lambda}_1),F_{2\mid\mathbf{X}}(y^{(2)}\mid\mathbf{x};\bm{\lambda}_2); \bm{\psi}\big), \nonumber
\end{align}
where $\bm{\theta} = (\bm{\lambda}_1,\bm{\lambda}_2,\bm{\psi})$, $f_\ell(\cdot;\bm{\lambda}_\ell)$ and $F_\ell(\cdot;\bm{\lambda}_\ell)$ denote the marginal density and CDF of $Y^{(\ell)}$, respectively, and $c(\cdot\,;\bm{\psi})$ is the copula density, which absorbs the entire dependence structure. Under the ``simplifying assumption'', conditioning on $\mathbf{X}$ acts exclusively through the conditional margins. We refer the interested reader to \cite{hobaek_haff_simplified_2010}, \cite{gijbels_estimation_2015}, \cite{grazian_approximate_2022}, and \cite{levi_bayesian_2018} for a thorough discussion of the simplifying assumption and its validity.

Compared to the standard multivariate parametric model (e.g., the multivariate ADAPTIVE is Section~\ref{sec: mvt_Greci}), this representation has the key advantage of flexibly decoupling the marginal behavior of each biomarker from their joint dependence structure. This considerably empowers model specification, since a wide range of parametric or semiparametric families can be entertained for each margin independently, while possibly simplifying posterior computation, as discussed in Section~\ref{sec: Bayes_IFM}.

Using Eq.~\eqref{eq: cond_cop}, we now generalize the Bayesian ADAPTIVE model described in Section~\ref{sec: uvt} to a flexible multivariate version as
\begin{align}\label{eq: hier_mod_mvt}
    Y_{i,j}^{(\ell)} \mid \bm{\beta}^{(\ell)}, \bm{\eta}^{(\ell)}  &\overset{ind}{\sim} \mathcal{D}^{(\ell)}\big(\mathbf{X}_{i,j}' \bm{\beta}^{(\ell)}, \bm{\eta}^{(\ell)}\big), \nonumber\\
    \bm{\beta}^{(\ell)} \mid \bar{\bm{\beta}}^{(\ell)}, \bar{\Sigma}_{\bm{\beta}^{(\ell)}} &\sim \mathcal{N}_p\big(\bar{\bm{\beta}}^{(\ell)}, \bar{\Sigma}_{\bm{\beta}^{(\ell)}}\big),\\
    \bm{\eta}^{(\ell)} \sim \pi_{\bm{\bar{\eta}}^{(\ell)}}, \nonumber
\end{align}
where $\mathcal{D}^{(\ell)}$ denotes a generic parametric family for margin $\ell$, with mean parameter $\mu^{(\ell)}_{i,j} \doteq \mathbf{X}_{i,j}'\bm{\beta}^{(\ell)}$, where $\bm{\beta}^{(\ell)}$ are the population-level (fixed) effects, and remaining (family-specific) parameters $\bm{\eta}^{(\ell)}$. Remaining parameters $\bar{\bm{\beta}}^{(\ell)}, \bar{\Sigma}_{\bm{\beta}^{(\ell)}},\bm{\bar{\eta}}^{(\ell)}$ represent prior hyperparameters and can be determined according to any historical population-level information or kept weakly/non informative. The hierarchical model in \eqref{eq: hier_mod_mvt} is completed by a parametric copula family $C_{\bm{\psi}}$ linking the conditional marginals
\begin{align}\label{eq: par_cop}
    \mathbf{U}_{i,j} \mid \bm{\psi} \overset{ind}{\sim} C_{\bm{\psi}}, \qquad \bm{\psi} \sim \pi_{\bm{\psi}},
\end{align}
where $\mathbf{U}_{i,j} \doteq \big(U^{(1)}_{i,j},  U^{(2)}_{i,j}\big)$, with $U^{(\ell)}_{i,j} \doteq F_{\ell\mid\mathbf{X}}\big(y^{(\ell)}_{i,j}\mid \mathbf{x}_{i,j}; \bm{\lambda}_\ell\big)$ by construction, and $\pi_{\bm{\psi}}$ is a prior on the copula parameter. 

\section{Bayesian computation in copula-based models}
\label{sec: Bayes_comp}

Bayesian inference for copula-based models of the form described in Section~\ref{sec: copula_mvt} poses non-trivial computational challenges, arising primarily from the need to {\it jointly} estimate the marginal parameters $\bm{\lambda}$ and the copula parameters $\bm{\psi}$ while properly propagating the uncertainty between the two components of the model. As discussed in Section~\ref{sec: Intro}, the classical solution to this problem is the IFM estimator of \cite{joe1996estimation, joe2005asymptotic}, conceived under a frequentist framework. This two-step strategy, whereby marginal parameters are estimated first and copula parameters are subsequently estimated conditional on these marginal estimates (all via MLE), has been adopted in applied financial contexts in \cite{hurlimann_fitting_2004} and \cite{roch_testing_2006}. A fully joint likelihood-based approach, optimizing simultaneously over $\bm{\theta} = (\bm{\lambda}, \bm{\psi})$, would be a valid 
alternative. However, this approach may suffer from slow mixing and scalability issues as the dimension $d$ grows, or when the marginal or copula densities are complex. From a Bayesian standpoint, this issue translates into a joint posterior distribution over $\bm{\theta}$ whose full conditional distributions rarely admit closed-form expressions, owing to the composite structure of the copula density underlying \eqref{eq: sklar1}. Since the combination of different copula and marginal densities typically lacks a standard form, the resulting posteriors are analytically intractable, so that tailored MCMC schemes, or more sophisticated approaches are required to carry out inference.

Most of the existing literature has focused on the Bayesian estimation of the copula component alone, treating the margins as fixed or pre-transformed; we refer to \cite{damien_bayesian_2013, shemyakin2017introduction} for a general treatment and to \cite{min2010bayesian, levi_bayesian_2018} for more specific cases. Silva and Lopes \cite{silva_copula_2008} were among the
first to explicitly explore joint copula-marginal estimation, proposing a full slice-sampling-based MCMC scheme for simple i.i.d. parametric families (no linear models). Through simulation studies, they showed that this joint approach better captures the posterior dependence, which in turn improves copula-based model selection. In a related direction, \cite{craiu2012mixed} extended joint Bayesian estimation to the conditional copula setting, where the copula parameter itself varies with a covariate, and confirmed a better understanding of the parameter dependences and improved model selection. 

Subsequent work has extended this joint-estimation logic in several contexts. In \cite{smith_modelling_2012} a joint MCMC strategy is adopted for the specific skew-$t$ copula, exploiting the conditionally Gaussian representation of the skew-$t$ distribution. Building on the Gaussian-copula data-augmentation scheme of \cite{pitt_efficient_2006}, in 
\cite{smith_estimation_2012} two data-augmentation MCMC schemes are proposed for copulas with discrete or mixed margins.  
More recently, \cite{haschka2026bayesian} develop a joint MCMC scheme for the joint sampling of regression coefficients and copula correlation in a Gaussian-copula endogeneity-correction model, underscoring the extended relevance of joint posterior estimation.

Building on this literature, and motivated by the trade-off it highlights between computational simplicity and full uncertainty propagation, we outline and compare two computational solutions for the multivariate copula-based ADAPTIVE model introduced in Section~\ref{sec: copula_mvt}. The first, proposed in \cite{deliu2025multivariatecopulabasedbayesianframework} extends the classical IFM logic to a Bayesian framework: marginal and copula parameters are both estimated in a Bayesian fashion, through a two-step (rather than integrated), but fully probabilistic, procedure. This is the object of Section~\ref{sec: Bayes_IFM}. In Section~\ref{sec: full_MCMC}, we instead present a novel joint MCMC alternative, that updates marginal and copula parameters within a single Markov chain, tailored specifically to our copula-based multivariate \textit{ADAPTIVE}. The two approaches are then compared empirically in Sections~\ref{sec: simulations}-\ref{sec: application}.

\subsection{Bayesian IFM} \label{sec: Bayes_IFM}

The classical IFM procedure adopt a frequentist approach to independently estimate the marginal parameters $\bm{\lambda}$ first (Step 1), and then, keeping the estimates fixed, a frequentist procedure is used to estimate $\bm{\psi}$. Such an approach is therefore likely to produce overly confident interval estimates by neglecting the sampling variability of $\hat{\bm{\lambda}}$ (as hinted in Section~\ref{sec: Intro}). To overcome this limitation, we proposed a Bayesian version of the IFM approach in \cite{deliu2025multivariatecopulabasedbayesianframework}. Marginal and copula parameters are estimated through the same two-step logic underlying the classical IFM procedure, but full Bayesian principles are adopted at \emph{both} steps, so that uncertainty is propagated from the marginal to the copula estimation stage.

Formally, recall $\bm{\theta} = (\bm{\lambda}_1,\bm{\lambda}_2,\bm{\psi})$, with $\bm{\lambda}_\ell$ the parameter vector of margin $\ell=1,2$, and $\bm{\psi}$ the copula parameter(s). Given a sample of size $n$, $\mathcal{F}_n = \{\mathbf{x}_i, \underline{y}_i\}_{i=1,\dots,n}$, the Bayesian IFM proceeds in two steps:
\begin{description}
    \item[\textbf{Step 1: Bayesian marginal estimation.}] For each margin $\ell = 1,2$, we obtain, independently of each other and of the copula component, a set of $M$ posterior draws $(\tilde{\bm{\lambda}}_{\ell,(1)}, \dots, \tilde{\bm{\lambda}}_{\ell,(M)})$ approximating the marginal posterior $\pi(\bm{\lambda}_\ell \mid \mathcal{F}_n)$, obtained via posterior sampling, e.g., MCMC methods~\citep{robert_monte_2004, hoffman2014no}.
    \item[\textbf{Step 2: Bayesian copula estimation.}] For each posterior vector draw $\tilde{\bm{\lambda}}_{(m)} = (\tilde{\bm{\lambda}}_{1,(m)}, \tilde{\bm{\lambda}}_{2,(m)})$, for $m = 1,\dots,M$, obtained in Step 1, pseudo-observations are generated as 
\begin{equation}\label{eq: pit}
    \tilde{u}_{i\ell,(m)} = F_\ell\big(y_{i\ell} \mid \mathbf{x}_i, \tilde{\bm{\lambda}}_{\ell,(m)}\big), \quad \ell = 1,2;\ i=1,\dots,n,
\end{equation}
so that $\tilde{\mathbf{u}}_{i,(m)} = (\tilde{u}_{i1,(m)}, \tilde{u}_{i2,(m)})$ represents a pseudo-sample from the copula, conditional on the $m$-th marginal draws. Then, a chain of length $M'$ of posterior draws is obtained for the copula parameter for each of the $m$-th marginal draws. The last draw $\tilde{\bm{\psi}}_{(m)} \doteq \tilde{\bm{\psi}}_{(M',m)}$ can be retained from the copula chain, assuming it has converged by iteration $M'$.
\end{description}
Repeating Step 2 across $m=1,\dots,M$ marginal draws yields a set of (pseudo-)joint posterior draws $\big(\tilde{\bm{\lambda}}_{1,(m)}, \tilde{\bm{\lambda}}_{2,(m)}, \tilde{\bm{\psi}}_{(m)}\big)$, $m=1,\dots,M$, which approximate the joint uncertainty in $\bm{\theta} = (\bm{\lambda}_1,\bm{\lambda}_2,\bm{\psi})$ and can be used directly to derive the predictive distributions of interest. The pseudo-code is summarized in Algorithm~\ref{alg: BIFM}. 
\begin{algorithm}
\caption{\enskip Pseudo-code for the Bayesian IFM}\label{alg: BIFM}
\begin{algorithmic}
  \State \textbf{Input:} Sample $\mathcal{F}_n = \{\mathbf{x}_i, \mathbf{y}_i\}_{i=1,\dots,n}$, $\mathbf{y}_i \in \mathbb{R}^2$, $\mathbf{x}_i \in \mathbb{R}^p$
  \State \phantom{\textbf{Input:}} Marginal and copula models, with associated parameters $(\bm{\lambda}_1, \bm{\lambda}_2, \bm{\psi})$
  \State \phantom{\textbf{Input:}} length of posterior draws $M, M' \in \mathbb{R}$
  \State \phantom{\textbf{Input:}} Priors
  on $(\bm{\lambda}_1, \bm{\lambda}_2, \bm{\psi})$
  \State \textbf{Output:} Posterior draws: $\{\big(\tilde{\bm{\lambda}}_{1,(m)}, \tilde{\bm{\lambda}}_{2,(m)}, \tilde{\bm{\psi}}_{(m)}\big)\}_{m=1,\dots,M}$
  \State
  \State Step 1: marginal parameters
  \For{$\ell = 1, 2$} 
  \State get $M$ posterior draws $(\tilde{\bm{\lambda}}_{\ell,(1)}, \dots, \tilde{\bm{\lambda}}_{\ell,(M)})$ 
approximating the marginal posterior distributions $\pi(\bm{\lambda}_\ell \mid \mathcal{F}_n)$
  \State generate pseudo-data
  \[
    \begin{pmatrix}
      \tilde{u}^{(\ell)}_{1,(1)} & \dots & \tilde{u}^{(\ell)}_{1,(M)} \\
      \vdots & \dots & \vdots \\
      \tilde{u}^{(\ell)}_{n,(1)} & \dots & \tilde{u}^{(\ell)}_{n,(M)}
    \end{pmatrix},
  \]
with $\tilde{u}^{(\ell)}_{i,(m)} \doteq F_\ell\big(y_{i\ell} \mid \tilde{\bm{\lambda}}_{\ell,(m)}, \mathbf{x}_i\big)$.
  \EndFor
  \State
  \State \textbf{Step 2: copula parameters}
  \For{$m = 1,\dots,M$}
  \State fix the marginal parameters to $(\tilde{\bm{\lambda}}_{1,(m)}, \tilde{\bm{\lambda}}_{2,(m)})$ and take associated pseudo-data $\{(\tilde{u}^{(1)}_{i,(m)}, \tilde{u}^{(2)}_{i,(m)})\}_{i=1,\dots,n}$
  \State get $M'$ posterior draws $(\tilde{\bm{\psi}}_{(1,m)}, \dots, \tilde{\bm{\psi}}_{(M',m)})$ approximating the copula posterior distribution
  \State set $\tilde{\bm{\psi}}_{(m)} \doteq \tilde{\bm{\psi}}_{(M',m)}$, assuming the copula chain has converged at step $M'$
  \EndFor
  \State
  \State \textbf{Return} posterior draws: $\{\big(\tilde{\bm{\lambda}}_{1,(m)}, \tilde{\bm{\lambda}}_{2,(m)}, \tilde{\bm{\psi}}_{(m)}\big)\}_{m=1,\dots,M}$
\end{algorithmic}
\end{algorithm}

Compared to its frequentist counterpart \citep{joe1996estimation}, the Bayesian IFM allows uncertainty to flow from the marginal to the copula parameters, providing a more reliable quantification of the overall estimation uncertainty, while retaining the computational simplicity of a two-step procedure. It can be recognized as an instance of the broader class of \emph{modular}, or \emph{cut}, Bayesian inference schemes \citep{jacob2017better, pompe2025asymptotics}, in which a model is partitioned into sequentially estimated modules (here, the marginal and copula components) with information deliberately prevented from flowing back from the second module into the first. This modular perspective has recently been studied for copula models independently by \cite{smith2025cutting}. Because our Bayesian IFM coincides with the type-1 cut posterior (second module represented by the copula and first module given by the marginals) of \cite{smith2025cutting}, it inherits their Lemma 1 (asymptotic equivalence to frequentist IFM), Lemma 2 (asymptotically correct uncertainty quantification for the marginal parameters even under copula misspecification), and Lemma 3 (asymptotically correct uncertainty quantification for the copula parameter under correct specification of both marginals and copula model). Nevertheless, since cutting discards the feedback from the copula to the marginal parameters, which is exploited by a joint posterior sampling, the Bayesian IFM and joint MCMC scheme may differ in their efficiency and shall coincide only when the marginal and copula parameters are asymptotically information-orthogonal. This can be understood as a direct consequence of comparing Lemma 3's variance expression (Bayesian IFM) to the block-inverse decomposition of the joint information matrix (see Supplementary material of \cite{smith2025cutting}). Since this may not be true in general, the development and comparison with a fully joint alternative remain of interest and it helps assess the impact of model misspecification in one of the modules.

\subsection{Joint MCMC} \label{sec: full_MCMC}

We now illustrate the proposed \emph{fully joint} Markov chain Monte Carlo (joint MCMC) algorithm that directly targets the posterior:
\begin{multline}
\pi(\bm \theta \mid \mathcal{F}_n)
\propto
\left[
\prod_{i=1}^n \prod_{j=1}^{t_i}
f_1(y_{i,j}^{(1)} \mid X_{i,j}; \bm \lambda_1)\,
f_2(y_{i,j}^{(2)} \mid X_{i,j}; \bm \lambda_2)\, \right .\\
\times \left .\,c\!\left(
F_1(y_{i,j}^{(1)}\mid X_{i,j};\bm \lambda_1),
F_2(y_{i,j}^{(2)}\mid X_{i,j};\bm \lambda_2); \bm \psi
\right)
\right] 
\pi(\bm \lambda_1)\,\pi(\bm \lambda_2)\,\pi(\bm \psi),
\end{multline}
with priors specified in (4)--(5). The composite structure of the likelihood prevents closed-form full conditionals, 
necessitating a Metropolis--Hastings (MH) strategy, outlined below.

Recall from Section~\ref{sec: Bayes_IFM} that $\bm{\theta} = (\bm{\lambda}_1,\bm{\lambda}_2,\bm{\psi})$. We design a blocked MH sampler of the form
\[
\bm \theta_{(m+1)} = 
\left(
\bm{\lambda}_{1, (m+1)},\,
\bm{\lambda}_{2, (m+1)},\,
\bm{\psi}_{(m+1)}
\right),\quad m=1,\dots,M,
\]
with $M$ the length of the posterior chain. 
For each margin $\ell = 1,2$, the parameter vector $\bm \lambda_\ell$ may include regression coefficients and scale parameters, while $\bm{\psi}$ relates to the copula parameters. 

\begin{description}
    \item[{\bf Marginal parameter updates.}] We use adaptive random-walk MH proposals of the form
\[
\bm \lambda_\ell^\ast = \bm \lambda_{\ell, (m)} + \bm \epsilon_{\ell, (m)},\qquad 
\bm \epsilon_{\ell, (m)} \sim \mathcal{N}(0,\,\Sigma_{\ell, (m)}),\quad \ell = 1,2.
\]
where $\Sigma_{\ell, (m)}$ is updated sequentially using the empirical covariance of the chain. 
The acceptance probability is
$$\alpha_\ell = 
\min \left\{
1,\,
\frac{
\pi(\bm \lambda_\ell^\ast,\bm \lambda_{-\ell,(m)},\bm \psi_{(m)}\mid \mathcal{F}_n)
}{
\pi(\bm \lambda_{\ell, (m)},\bm \lambda_{-\ell,(m)},\bm \psi_{(m)}\mid \mathcal{F}_n)
} \right \},\quad \ell = 1,2.$$
Because the copula density depends on the transformed variables $U_{i,j}^{(\ell)} = F_\ell(y_{i,j}^{(\ell)}\mid X_{i,j};\bm \lambda_\ell)$, each proposed $\bm \lambda_\ell^\ast$ requires recomputing the pseudo-observations entering the copula likelihood. This ensures that dependence information influences the marginal updates, unlike in the Bayesian IFM, where Step~1 is performed independently of Step-2. 
\item[{\bf Copula parameter update.}] For the dependence parameter $\bm \psi$, we use an adaptive MH step
\[
\bm \psi^\ast = \bm \psi_{(m)} + \bm \eta_{(m)},\quad 
\bm \eta_{(m)} \sim \mathcal{N}(0,\sigma^2_{\bm \psi,(m)}),
\]
with acceptance probability
\[
\alpha_{\bm\psi} = 
\min\left\{
1,\,
\frac{
\prod_{i,j}
c\!\left(
U_{i,j}^{(1)}, U_{i,j}^{(2)}; \bm \psi^\ast
\right)
\pi(\bm \psi^\ast)
}{
\prod_{i,j}
c\!\left(
U_{i,j}^{(1)}, U_{i,j}^{(2)}; \bm \psi_{(m)}
\right)
\pi(\bm \psi_{(m)})
}
\right\}.
\]
This update is computationally demanding because the copula must be evaluated for all observations at each iteration. 
\end{description}
The pseudo-code is summarized in Algorithm~\ref{alg: FullMCMC}. 

Compared to the Bayesian IFM procedure of Section~\ref{sec: Bayes_IFM}, the joint MCMC
algorithm provides a coherent and fully joint exploration of the posterior
distribution of $\bm \theta$, thereby avoiding potential underestimation of uncertainty inherent to two-stage approaches. By updating all components of $\bm \theta$ within a single Markov chain, the method correctly propagates dependence between marginal and copula parameters, yielding posterior intervals that appropriately reflect joint variability.

\begin{algorithm}
\caption{\enskip Pseudo-code for the joint MCMC}\label{alg: FullMCMC}
\begin{algorithmic}
  \State \textbf{Input:} Sample $\mathcal{F}_n = \{\mathbf{x}_i, \mathbf{y}_i\}_{i=1,\dots,n}$, 
        $\mathbf{y}_i \in \mathbb{R}^2$, $\mathbf{x}_i \in \mathbb{R}^p$
  \State \phantom{\textbf{Input:}} Marginal and copula models, with associated parameters 
        $(\bm{\lambda}_1, \bm{\lambda}_2, \bm{\psi})$
  \State \phantom{\textbf{Input:}} length of posterior draws $M \in \mathbb{R}$
  \State \phantom{\textbf{Input:}} Priors
  on $(\bm{\lambda}_1, \bm{\lambda}_2, \bm{\psi})$
  \State \textbf{Output:} Posterior draws: 
        $\{\big(\bm{\lambda}_{1,(m)}, \bm{\lambda}_{2,(m)}, \bm{\psi}_{(m)}\big)\}_{m=1,\dots,M}$
  \State
  \State Initialize $(\bm{\lambda}_{1,(0)}, \bm{\lambda}_{2,(0)}, \bm{\psi}_{(0)})$
  \State
  \For{$m = 1,\dots,M$}
      \State
      \State \textbf{Step 1: update marginal parameters for Hgb}
      \State Propose $\bm{\lambda}_1^\ast = \bm{\lambda}_{1,(m-1)} + \epsilon_1$, 
             with $\epsilon_1 \sim \mathcal{N}(0, \Sigma_1)$
      \State Compute acceptance probability
      
\[
      \alpha_1 = 
      \min\left\{
      1,\,
      \frac{
      \pi(\bm{\lambda}_1^\ast, \bm{\lambda}_{2,(m-1)}, \bm{\psi}_{(m-1)} \mid \mathcal{F}_n)
      }{
      \pi(\bm{\lambda}_{1,(m-1)}, \bm{\lambda}_{2,(m-1)}, \bm{\psi}_{(m-1)} \mid \mathcal{F}_n)
      }
      \right\}
      \]
      \State Set $\bm{\lambda}_{1,(m)} = \bm{\lambda}_1^\ast$ with probability $\alpha_1$, 
             otherwise $\bm{\lambda}_{1,(m)} = \bm{\lambda}_{1,(m-1)}$
      \State
      \State \textbf{Step 2: update marginal parameters for OFFs}
      \State Propose $\bm{\lambda}_2^\ast = \bm{\lambda}_{2,(m-1)} + \epsilon_2$, 
             with $\epsilon_2 \sim \mathcal{N}(0, \Sigma_2)$
      \State Compute acceptance probability
      
\[
      \alpha_2 = 
      \min\left\{
      1,\,
      \frac{
      \pi(\bm{\lambda}_{1,(m)}, \bm{\lambda}_2^\ast, \bm{\psi}_{(m-1)} \mid \mathcal{F}_n)
      }{
      \pi(\bm{\lambda}_{1,(m)}, \bm{\lambda}_{2,(m-1)}, \bm{\psi}_{(m-1)} \mid \mathcal{F}_n)
      }
      \right\}
      \]
      \State Set $\bm{\lambda}_{2,(m)} = \bm{\lambda}_2^\ast$ with probability $\alpha_2$, 
             otherwise $\bm{\lambda}_{2,(m)} = \bm{\lambda}_{2,(m-1)}$
      \State
      \State \textbf{Step 3: update copula parameter}
      \State Propose $\bm{\psi}^\ast = \bm{\psi}_{(m-1)} + \eta$, 
             with $\eta \sim \mathcal{N}(0, \sigma_\psi^2)$
      \State Compute acceptance probability
      
\[
      \alpha_\psi = 
      \min\left\{
      1,\,
      \frac{
      \pi(\bm{\lambda}_{1,(m)}, \bm{\lambda}_{2,(m)}, \bm{\psi}^\ast \mid \mathcal{F}_n)
      }{
      \pi(\bm{\lambda}_{1,(m)}, \bm{\lambda}_{2,(m)}, \bm{\psi}_{(m-1)} \mid \mathcal{F}_n)
      }
      \right\}
      \]
      \State Set $\bm{\psi}_{(m)} = \bm{\psi}^\ast$ with probability $\alpha_\psi$, 
             otherwise $\bm{\psi}_{(m)} = \bm{\psi}_{(m-1)}$
      \State
      \State Store draw: 
            $\big(\bm{\lambda}_{1,(m)}, \bm{\lambda}_{2,(m)}, \bm{\psi}_{(m)}\big)$
  \EndFor
  \State
  \State \textbf{Return} posterior draws: 
        $\{\big(\bm{\lambda}_{1,(m)}, \bm{\lambda}_{2,(m)}, \bm{\psi}_{(m)}\big)\}_{m=1,\dots,M}$
\end{algorithmic}
\end{algorithm}

\section{Comparative Simulation Studies}\label{sec: simulations}
Simulation studies are first conducted to assess the inferential performance of the joint MCMC and Bayesian IFM approaches under controlled conditions. 
To provide a realistic benchmark for the subsequent real-world application, the data-generating process is calibrated to reproduce the main distributional and dependence features observed in the original data described in Section~\ref{sec: application}. 
Specifically, following the patterns shown in Figure~\ref{fig:data} and Table~\ref{tab:aic}, for each simulated dataset, we use Gaussian marginals combined with a Gumbel copula. Gaussian marginals were chosen to isolate the effects of copula specification and inferential methodology, while keeping the data-generating process simple. The copula itself is selected from four candidates - Gaussian, Clayton, Gumbel, and Frank - using the Akaike Information Criterion (AIC), see Table~\ref{tab:aic}. 

Among the available variables, sex and age are selected as the main covariates for the simulation study. The data-generating process (i.e., the marginal parameters) is guided by fitting a linear model to the observed data for each biomarker, with results reported in the Appendix~\ref{appendix: fit}. 
The copula density is assumed to be covariate independent, with the Gumbel dependence parameter estimated using the \texttt{BiCopEst()}~\citep{vinecopula} and resulting in a Kendall's $\tau \approx 0.55$. Specifically, for $i=1,\dots,n$ the generative process is given by
\begin{align}
    &Y_{i}^{\texttt{Hgb}} \sim \mathcal{N}(\mu_{i} = 14.18 + 1.54 \, X^{(1)}_{i} - 0.03\, X^{(2)}_{i},\, \sigma = 0.95), \\
       &Y_{i}^{\texttt{OFFs}} \sim \mathcal{N}(\mu_{i} = 51.50 + 22.29 \, X^{(1)}_{i} + 0.22\,   X^{(2)}_{i},\, \sigma = 12.99), \nonumber\\
    &(U_{i}^{\texttt{Hgb}}, U_{i}^{\texttt{OFFs}}) \sim C_\theta^\texttt{Gumbel} (\theta = 2.22), \nonumber
\end{align} 
where $X^{(1)}_{i}$ denotes the sex of athlete $i$, and $X^{(2)}_{i}$ denotes the corresponding age. To mimic the true variability of the covariates, rather than generating them ex-novo, we simulate them using a bootstrap resampling approach. We consider a total sample size of $n = 500$.

Once the observations are generated on the uniform scale from the copula, they are transformed back to the original scale using the inverse marginal CDF, with the marginal means evaluated conditionally on the corresponding covariate values, $Y_{i}^{(\ell)} = \Phi^{-1}_{(\ell)}(U^{(\ell)}; \mu_{i}^{(\ell)}, \sigma^{(\ell)})$.
Because the two biomarkers are measured on different scales, the simulated responses were standardized to improve numerical stability and facilitate prior specification. Since copulas are invariant under strictly increasing marginal transformations, this approach does not alter the dependence structure.
Weakly informative priors are assigned to the marginal parameters. Specifically, given $\mu_{i}^{(\ell)} = \beta_0^{(\ell)} + \beta_1^{(\ell)} \, X^{(1)}_{i} + \beta_2^{(\ell)}  \, X^{(2)}_{i}$, we set
\begin{align}
    \beta_k^{(\ell)} &\sim \mathcal{N}(0, 2.5^2),\quad \ell = 1,2; \\
    \sigma^{(\ell)} &\sim \mathrm{half}-t_3(0, 2.5), \quad \ell = 1,2,\nonumber
\end{align}
for all fixed coefficients $k = 0,1,2$.

Dependence is modeled directly through the parameter $\theta$ of the Gumbel copula, controlling the positive upper-tail dependence between the two biomarkers. Since only positive dependence is considered, a weakly informative uniform prior bounded in $(1,10)$ is assigned; that is,
\begin{align}
    (U_{i}^{\texttt{Hgb}}, U_{i}^{\texttt{OFFs}}) | \theta &\overset{ind}{\sim} C^\texttt{Gumbel}_\theta, \quad \theta \sim \mathcal{U}(1,10). \nonumber
\end{align}

All simulation studies are based on a number of $1{,}000$ Monte Carlo (MC) replicas.

\subsection{Software and computational details}
The inference was carried out under identical computational settings for both approaches. 
For each method, $4$ independent Markov chains were run, each consisting of $M=10{,}000$ iterations. The first half was discarded as burn-in, resulting in $5{,}000$ iterations. 
A thinning factor of 5 was subsequently applied, yielding $1{,}000$ posterior draws per chain and $4{,}000$ posterior draws overall.
All analyses were performed in \texttt{R} version 4.5.1 (2025-06-13). Bayesian inference for the marginal distributions was carried out using \texttt{rstan}~\cite{rstan} and the \texttt{brms}~\cite{brms} package, an interface for fitting Bayesian regression models with Stan.
The joint MCMC approach was performed using user-defined functions implementing random-walk Metropolis Hastings (RWMH) algorithms for the parameters of the two variables considered. 
Inference for the copula component was conducted analogously. Specifically, for both approaches, a user-defined RWMH function 
was used to sample from the posterior distribution of the dependence parameter of the selected copula family.
Copula selection was performed using the \texttt{BiCopSelect} function from the \texttt{VineCopula}~\citep{vinecopula} package. 
Finally, the computational cost was assessed using \texttt{Sys.time}.

\subsection{Comparative Results}
\begin{figure}[!htbp]
    \centering
    \includegraphics[width=\linewidth]{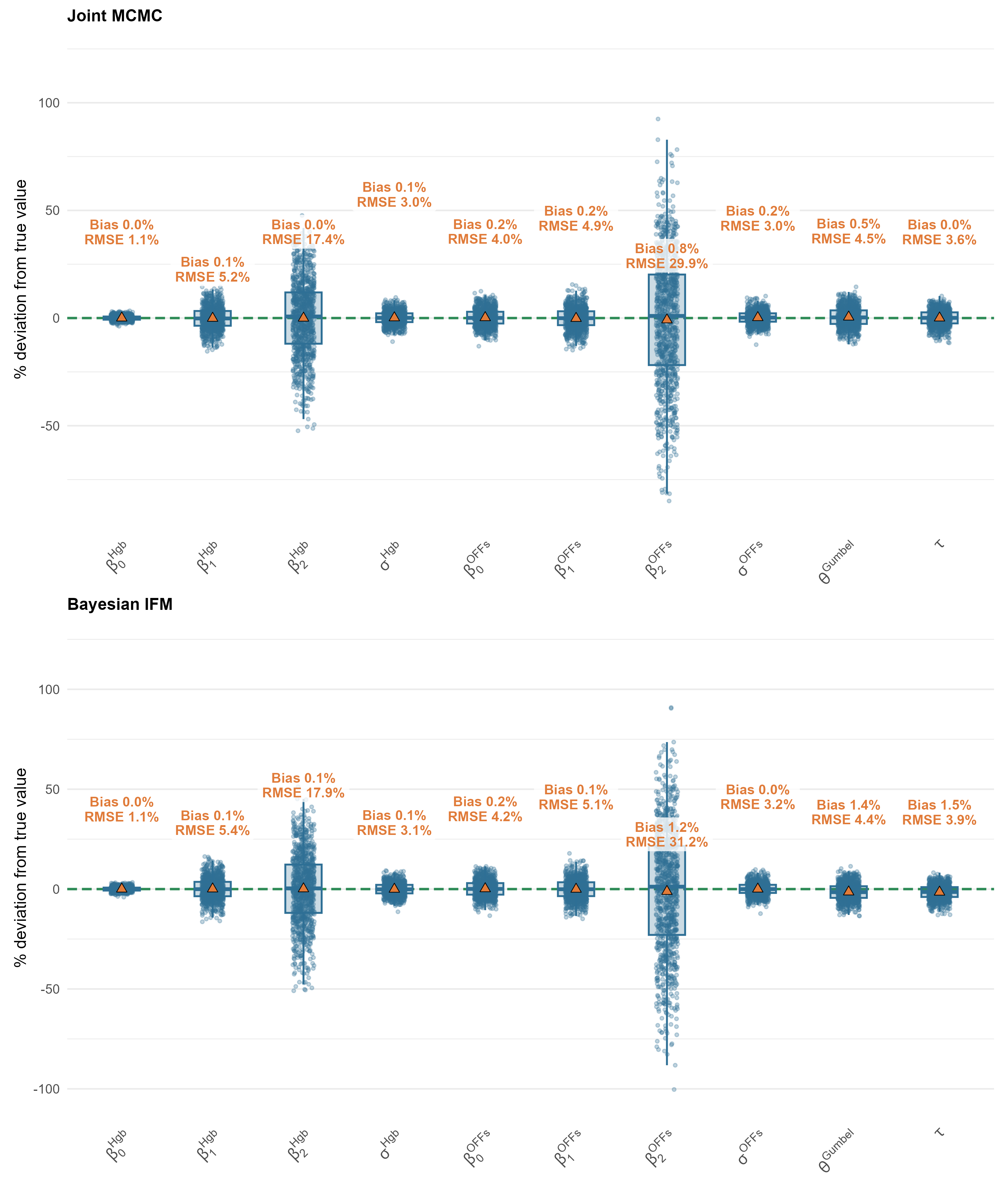}
    \caption{Boxplots show the percentage deviation of each simulation's posterior mean from the true parameter value, $100\times(\hat\theta-\theta)/|\theta|$, across 1000 simulated datasets of size $n=500$ under the Gumbel copula with covariates. Top: joint MCMC estimator. Bottom: Bayesian IFM. The orange triangle marks the mean relative bias per parameter, and the dashed line at zero marks an unbiased estimate.}
    \label{fig:deviation_cov}
\end{figure}

Figure~\ref{fig:deviation_cov} shows the distribution of the parameter estimates obtained from the two approaches across the $1{,}000$ Monte Carlo replications under the Gumbel copula specification (well-specified copula model). Both estimators show small bias across all parameters -as indicated by the deviation from the true value being close to zero- with $\beta_2$, associated with the Age effect, showing the widest spread and correspondingly higher $RMSE$ under both methods. The most notable differences are observed for the parameters related to copula and Kendall’s $\tau$, for which joint MCMC exhibits a lower bias than Bayesian IFM, while the RMSE values remain essentially identical between the two approaches.
Despite this variability, the distributions remain centered around the true value, indicating that the greater dispersion is not accompanied by a systematic bias.
Substantial differences emerge in terms of computational efficiency. For a simulated dataset with $n=500$ observations, obtaining the posterior samples required approximately 17.04 seconds with the joint MCMC, compared with approximately 159.03 seconds (2:39 minutes) with the Bayesian IFM. This corresponds to approximately 3.52 and 0.38 complete runs per minute, respectively. Thus, the joint MCMC is computationally more efficient, it completes approximately 9.33 times as many runs per minute as the Bayesian IFM.

\subsubsection{Copula misspecification}
To further evaluate the computational efficiency of the inferential methods under copula misspecification, three alternative copula families (Gaussian, survival Clayton, and Frank) are also fitted to the simulated data. Each of these copulas exhibits distinct shapes, capturing dependence in a different way.
As we are interested in positive dependence, weakly informative uniform priors are assigned to the respective dependence parameters, ensuring a support comparable
to the same range of Kendall's $\tau$ considered for the Gumbel copula, $(0, 0.9)$.
Parameter estimates are also reported in terms of Kendall's $\tau$, which provides a copula-invariant measure of dependence and therefore a common scale for comparison. This allows the analysis to focus on the effects of copula misspecification. 

\begin{figure*}[!htbp]
    \centering
    \includegraphics[width=1\linewidth]{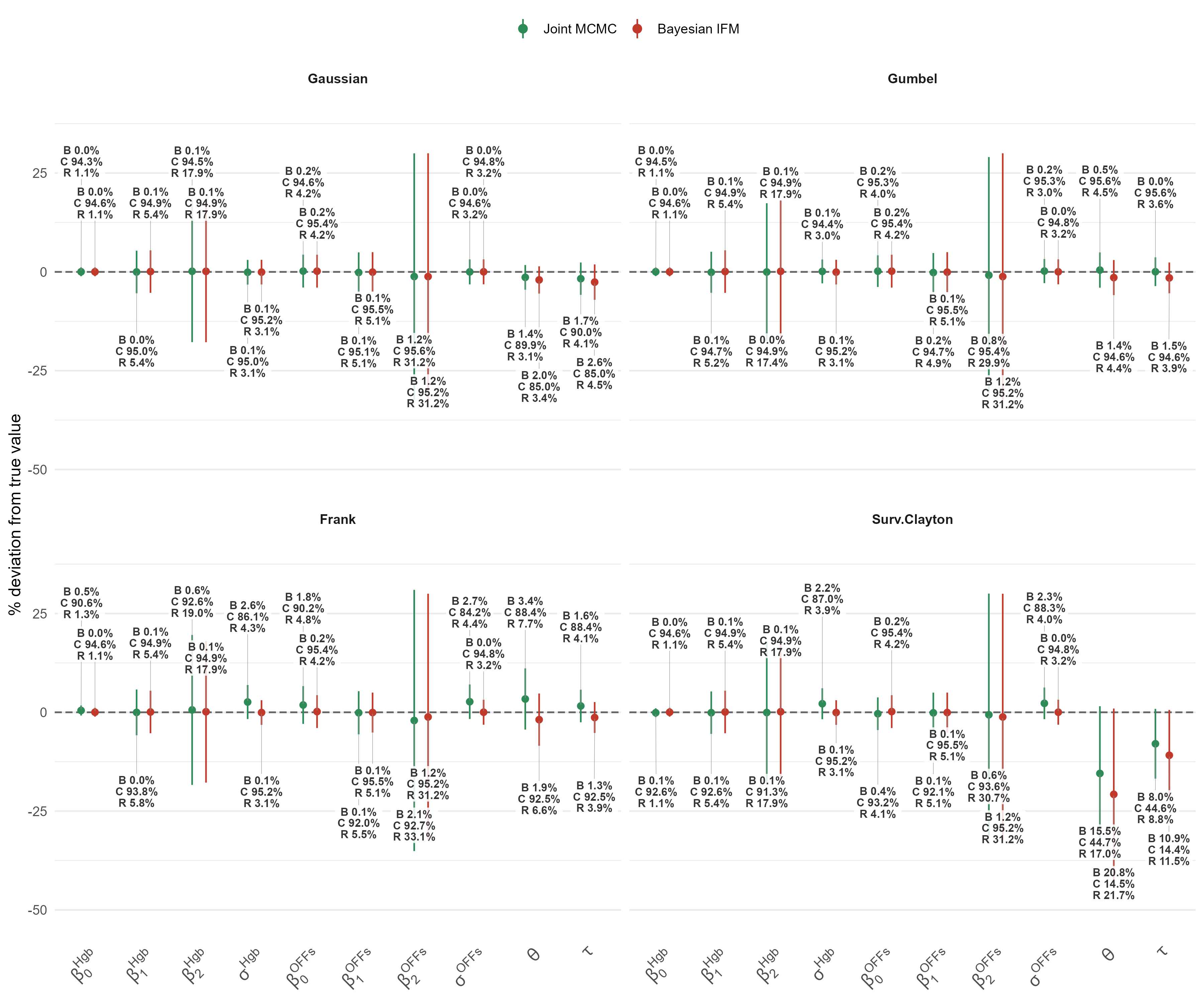}
    
    \caption{Points show the mean relative bias per parameter for the joint MCMC (green) and Bayesian IFM (red) estimators, under the Gaussian (top-left), Surv. Clayton (top-right), Gumbel (bottom-left) and Frank (bottom-right) copulas with covariates. Labels report relative bias (B), coverage of the 95\% credible interval (C), and relative RMSE (R).}
    \label{fig:comp_gumbel_frank}
\end{figure*}

Figure~\ref{fig:comp_gumbel_frank} shows the results obtained by applying the two approaches under the different copula models, reporting the relative bias, RMSE, and coverage probabilities for all parameters. The copula dependence parameter and $\tau$ show noticeably lower coverage under Frank than under Gumbel for both estimators, while the marginal parameters remain comparable across the two copula families.These findings are consistent with the theoretical results of \cite{smith2025cutting} on cutting feedback in misspecified copula models. Under correct specification of the marginal models, they establish that inference for the marginal parameters is asymptotically well-calibrated regardless of whether the copula function is correctly specified or not, while the dependence parameter requires both the marginals and the copula to be correctly specified. Our results reproduce this asymmetry. Coverage of the 95\% confidence intervals for the marginal parameters of Hgb and OFFs remained close to nominal across all four fitted copula families 
and under both the joint MCMC and Bayesian IFM estimators (0.842-0.956), indicating that misspecification of the copula has negligible impact on marginal inference when the margins themselves are correctly specified. In contrast, coverage for the dependence parameters ($\theta$ and $\tau$) was strongly sensitive to whether the fitted copula matched the true dependence structure. When the correctly specified copula was fit, coverage for $\theta$ and $\tau$ remained near nominal under both estimators (0.956/0.946, joint MCMC/Bayesian IFM). When the copula was misspecified, it deteriorated substantially, most severely for the Bayesian IFM estimator under the Clayton fit, where coverage for $\theta$ and $\tau$ fell to 0.145 and 0.144, respectively - compared to 0.447 and 0.446 under the joint MCMC estimator. This pattern mirrors the degradation in $\tau$ coverage reported by \cite{smith2025cutting} (Table 2) under copula misspecification, and supports their theoretical result that two-step (IFM-style) inference for the dependence parameter is particularly vulnerable to copula misspecification, since it does not propagate the additional uncertainty that a jointly estimated model can partially absorb. 

\section{Application to NADO Italia data}\label{sec: application}



The dataset used in the empirical analysis consists of longitudinal hematological measurements collected from athletes monitored through the Athlete Biological Passport (ABP) program. 
The dataset was provided by NADO Italia (\url{https://www.nadoitalia.it}), the sole authority responsible for implementing the anti-doping standards in Italy.


\begin{center}
\begin{table*}[!thb]%
\caption{Descriptive statistics of the study population by Age, Altitude, Test Type, and Sex, reported as absolute frequencies and relative percentages. \label{tab: covariates}}
\begin{tabular*}{\textwidth}{@{\extracolsep\fill}lcc|lcc|lcc|lcc|@{\extracolsep\fill}}
\toprule
& \multicolumn{2}{@{}c}{\textbf{Age}} & & \multicolumn{2}{@{}c}{\textbf{Altitude}} & & \multicolumn{2}{@{}c}{\textbf{Test Type}} & & \multicolumn{2}{@{}c}{\textbf{Sex}} \\\cmidrule{2-3}\cmidrule{5-6}\cmidrule{8-9}\cmidrule{11-12}
 & \textbf{Freq.}  & \multicolumn{1}{@{}c}{\textbf{Prop.}}  &   & \textbf{Freq.} & \multicolumn{1}{@{}c}{\textbf{Prop.}} &   & \textbf{Freq.} & \multicolumn{1}{@{}c}{\textbf{Prop.}} &   & \textbf{Freq.} & \multicolumn{1}{@{}c}{\textbf{Prop.}} \\
\midrule
$20\dashv35$ & $188$  & $85\%$  & No & $186$  & $83\%$ & Out-of-Comp  & $216$ & $97\%$ & M  & $113$ & $51\%$  \\
$35\dashv45$ & $14$  & $6\%$  & Yes & $31$  & $14\%$ & In-Comp  & $7$ & $3\%$ & F  & $110$ & $49\%$ \\
$45\dashv55$ & $21$  & $9\%$  & Unknow & $6$  & $3\%$ & & & \multicolumn{1}{@{}c}{} & & & \multicolumn{1}{@{}c}{} \\
\bottomrule
\end{tabular*}
\end{table*}
\end{center}

\subsection{Data description}
\begin{figure*}[!ht]
    \centering
    \includegraphics[width=\linewidth]{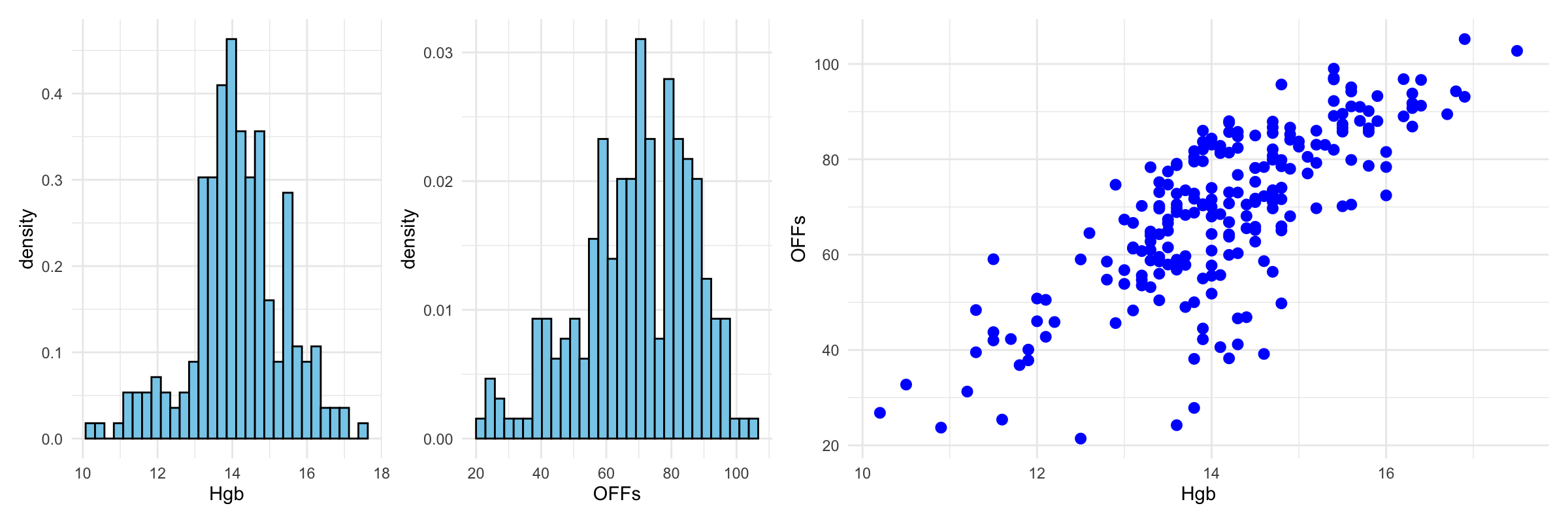}
    \caption{Histograms (left) and scatterplot (right) of the Hgb and OFFs markers, illustrating the marginal distributions and the presence of positive dependence.}
    \label{fig:data}
\end{figure*}



The dataset comprises the ABP profiles of a cohort of $n = 23$ athletes, collected between 2019 and 2024, for a total of $223$ repeated observations. For each athlete $i=1,\dots,n$, demographic and athlete-level information is available, along with all variables specified in the hematological module of the ABP used by WADA for the longitudinal monitoring of athletes.
Table~\ref{tab: covariates} shows the distribution of the main covariates considered. 
The dataset is well balanced with respect to sex, with $51\%$ of the observations relating to male athletes and the remaining $49\%$ relating to female athletes. Measurements are also concentrated mainly among athletes at the peak of their competitive careers: approximately $85\%$ of the observations relate to individuals aged between $20$ and $35$ years, while the remaining $15\%$ are distributed between $35$ and $55$ years of age.
By contrast, the covariates related to high-altitude exposure and the type of test show a highly imbalanced distribution. Only $24\%$ of the observations were collected after exposure to high altitudes, while the remainder of the sample comprises athletes who were not exposed or for whom no information is available. An even more pronounced imbalance is observed for the variable related to sample collection type, with only $3\%$ of the measurements taken in-competition.
Therefore, their contribution is not considered in the present study. 
The empirical distributions of the two primary biomarkers are shown in Fig.~\ref{fig:data}. 
Hgb shows relatively low variability, with observed values ranging approximately between $10$ g/dL and $18$ g/dL and its distribution exhibits slight negative skewness, with observations concentrated towards the upper end of the distribution.
The OFFs shows substantially greater variability, with values ranging approximately between $20$ and $100$. Its distribution also exhibits negative skewness, with a greater concentration of mass in the upper tail relative to Hgb.

\subsection{Model formulation}



\paragraph{Marginals}
Given the primary objective of comparing the joint MCMC and Bayesian IFM inferential procedures, and consistent with the ADAPTIVE model \citep{sottas_bayesian_2007}, Gaussian marginal distributions are assumed for both biomarkers.
\begin{align}
    Y_{i}^{(\ell)}| \beta^{(\ell)}, \sigma^{(\ell)} \sim \mathcal{N}(\mu_{i}^{(\ell)}, \sigma^{(\ell)}),\quad \ell = 1,2,
\end{align}
where $\mu_{i}^{(\ell)} = \beta_0^{(\ell)} + \beta_1^{(\ell)} \, X^{(1)}_{i} + \beta_2^{(\ell)}  \, X^{(2)}_{i}$.

Before fitting the model, both biomarkers are standardized using their respective empirical means and standard deviations. Relatively weakly informative priors are then assigned to the parameters of each marginal model 
\begin{align}
    \beta_k^{(\ell)} &\sim \mathcal{N}(0, 2.5^2),\quad \ell = 1,2; \\
    \sigma^{(\ell)} &\sim \mathrm{half}-t_3(0, 2.5), \quad \ell = 1,2,\nonumber
\end{align}
for all fixed coefficients $k = 0,1,2$.

\paragraph{Copula model}
The copula specification was selected based on the AIC criterion, reported in Table~\ref{tab:aic}.
The dependence structure between the biomarkers is modeled using a Gumbel copula, assigning a weakly informative prior to its dependence parameter
\begin{align}
    (U_{i}^{\texttt{Hgb}}, U_{i}^{\texttt{OFFs}}) | \theta \overset{ind}{\sim} C_\theta^{\texttt{Gumbel}}, \quad \theta \sim \mathcal{U}(1,10)
\end{align} 
allowing a broad range of positive dependence, from independence to strong upper-tail dependence.

\begin{table}[!htbp]
    \centering
    \begin{tabular*}{\linewidth}{@{\extracolsep\fill}lcc >{\columncolor{green!20}}c c@{\extracolsep\fill}}
    \toprule
        & \textsc{gaussian} & \textsc{clayton} & \textsc{gumbel} & \textsc{frank} \\ 
    \midrule
        \textbf{AIC} & -191.332 & -169.433 & -195.357 & -174.045 \\
        \textbf{BIC} & -187.924 & -166.026 & -191.950 & -170.638 \\
        \textbf{logLik} & 96.666 & 85.717 & 98.679 & 88.023 \\
    \bottomrule
    \end{tabular*}
    \caption{AIC, BIC, and log-likelihood values for the different copula specifications, evaluated on the NADO Italia dataset. The Gumbel copula provides the best fit according to all three criteria.}
    \label{tab:aic}
\end{table}

\subsubsection{Results and discussion}
Table~\ref{tab: res} reports the posterior point estimates and corresponding credible intervals obtained using the two approaches, which show substantial overlap. The main difference arises for the parameter $\beta_2^{\texttt{OFFs}}$, for which the joint MCMC credible interval includes zero, whereas the Bayesian IFM interval is entirely positive, suggesting stronger evidence for a positive effect. Overall, both approaches yield similar results. 
\begin{center}
\begin{table*}[!thb]%
\caption{Posterior estimates and 95\% credible intervals for the marginal regression coefficients, standard deviations, and Gumbel copula dependence parameters obtained from the NADO Italia dataset using the joint MCMC and Bayesian IFM approaches. The parameters $\beta_0^{(\ell)}$, $\beta_1^{(\ell)}$, and $\beta_2^{(\ell)}$ denote the intercept, sex effect, and age effect, respectively, for biomarker $\ell$. \label{tab: res}}
\begin{tabular*}{\textwidth}{@{\extracolsep\fill}lcccccc@{\extracolsep\fill}}
\toprule
& \multicolumn{3}{c}{{\bf joint MCMC}} & \multicolumn{3}{c}{{\bf Bayesian IFM}}\\ \cline{2-4}\cline{5-7} \addlinespace
& CI Upper & CI Lower & Estimate  & Estimate & CI Lower & CI Upper\\
\midrule
$\beta_0^{\texttt{Hgb}}$ & 14.789 & 13.816 & 14.320 & 14.180 & 13.721 & 14.646 \\
$\beta_1^{\texttt{Hgb}}$ & 1.725 & 1.253 & 1.489 & 1.534 & 1.281 & 1.782 \\
$\beta_2^{\texttt{Hgb}}$ & -0.016 & -0.046 & -0.031 & -0.028 & -0.042 & -0.014 \\
$\sigma^{\texttt{Hgb}}$ & 1.093 & 0.904 & 0.993 & 0.954 & 0.871 & 1.049 \\ \addlinespace
$\beta_0^{\texttt{OFFs}}$ & 61.265 & 48.794 & 55.153 & 51.562 & 45.292 & 57.833 \\
$\beta_1^{\texttt{OFFs}}$ & 23.564 & 17.132 & 20.348 & 22.243 & 18.710 & 25.658 \\
$\beta_2^{\texttt{OFFs}}$ & 0.335 & -0.058 & 0.136 & 0.216 & 0.023 & 0.408 \\ 
$\sigma^{\texttt{OFFs}}$ & 14.687 & 12.233 & 13.376 & 13.065 & 11.905 & 14.361 \\ \addlinespace
$\theta$ & 2.155 & 1.657 & 1.891 & 1.785 & 1.574 & 2.007 \\
$\tau$ & 0.536 & 0.396 & 0.469 & 0.437 & 0.365 & 0.502 \\
\bottomrule
\end{tabular*}
\end{table*}
\end{center}

\begin{figure}[!htbp]
    \centering
    \includegraphics[width=\linewidth]{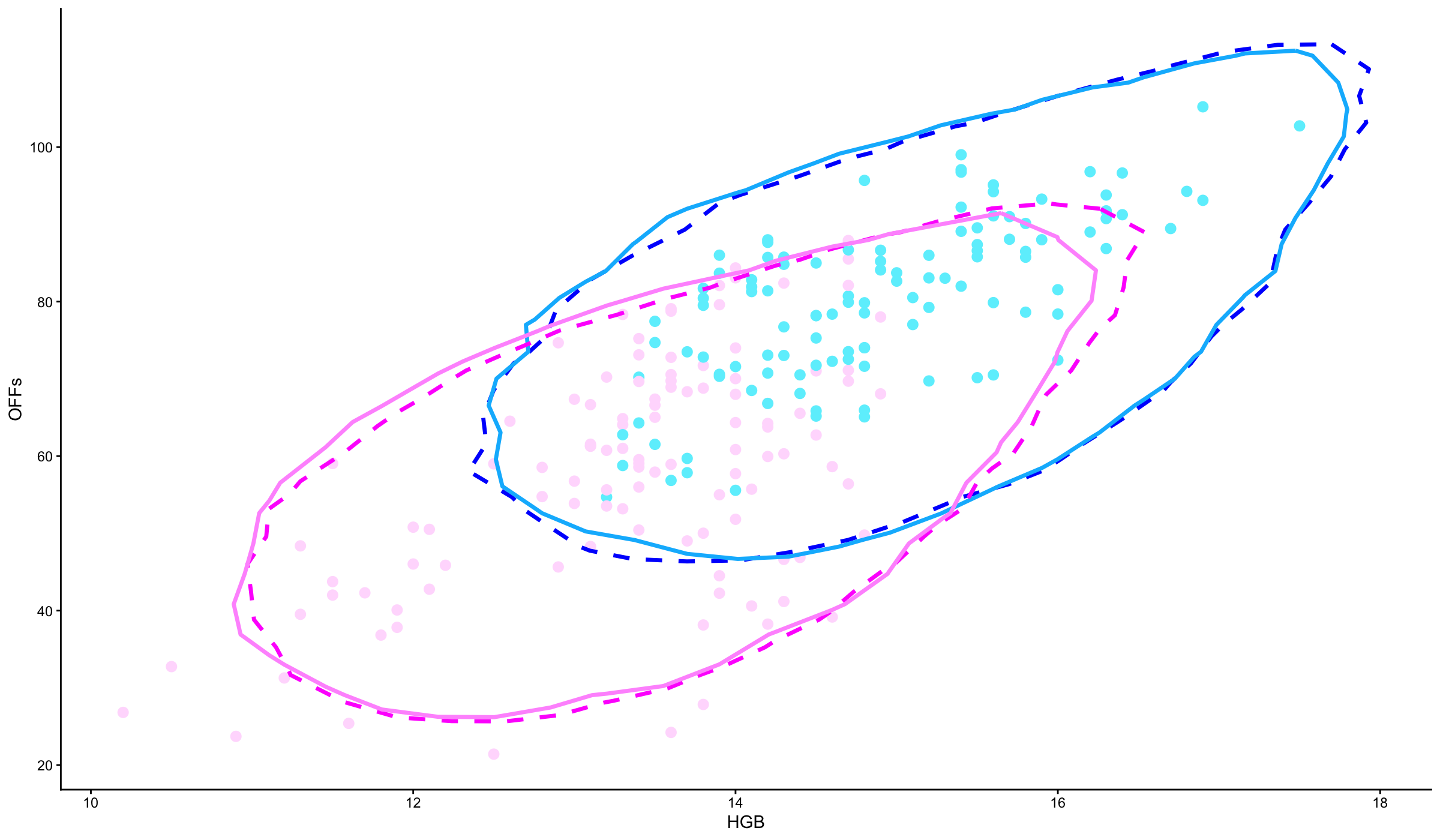}
    \caption{Estimated 95\% HPRs obtained with both the joint MCMC (dashed lines) and Bayesian IFM (solid lines) approaches. Points represent observations from the NADO Italia dataset. Blue and pink regions correspond to male and female athletes, respectively.}
    \label{fig: hpr}
\end{figure}

Figure~\ref{fig: hpr} shows the estimated 95\% HPRs for the NADO Italia dataset for both male and female athletes, with age fixed at 24 years, using both inferential approaches.
Male athletes exhibit higher values on average, resulting in a predictive region shifted towards the upper-right portion of the plot, whereas for female athletes the region is shifted towards the lower-left.
However, the regions exhibit a substantial area of overlap,
indicating the existence of a range values compatible with both sexes. 
All observations for male athletes fall within the respective HPRs under both inferential approaches, whereas a portion of the observations for female athletes falls outside. Specifically, 8.2\% and 9.2\% of female observations fall outside the HPRs estimated using the joint MCMC and Bayesian IFM approaches, respectively. These results are also summarized in Table~\ref{tab: fpr}.
\begin{table}[!htbp]
    \centering
    \begin{tabular*}{\linewidth}{@{\extracolsep\fill}lccc@{\extracolsep\fill}}
    \toprule
        & \multicolumn{3}{c}{\bf FPR} \\ \cline{2-4} \addlinespace
        & Male & Female & {\bf Total} \\ 
    \midrule\addlinespace
        \textbf{joint MCMC} & 0.000 & 0.082 & 0.040 \\ \addlinespace
        \textbf{Bayesian IFM} & 0.000 & 0.091 & 0.045 \\
    \bottomrule
    \end{tabular*}
    \caption{Classification performance of the predictive regions obtained from the joint MCMC and Bayesian IFM approaches in terms of false positive rate (FPR).}
    \label{tab: fpr}
\end{table}

\paragraph{Doped athletes}

\begin{figure}[!htbp]
    \centering
    \includegraphics[width=\linewidth]{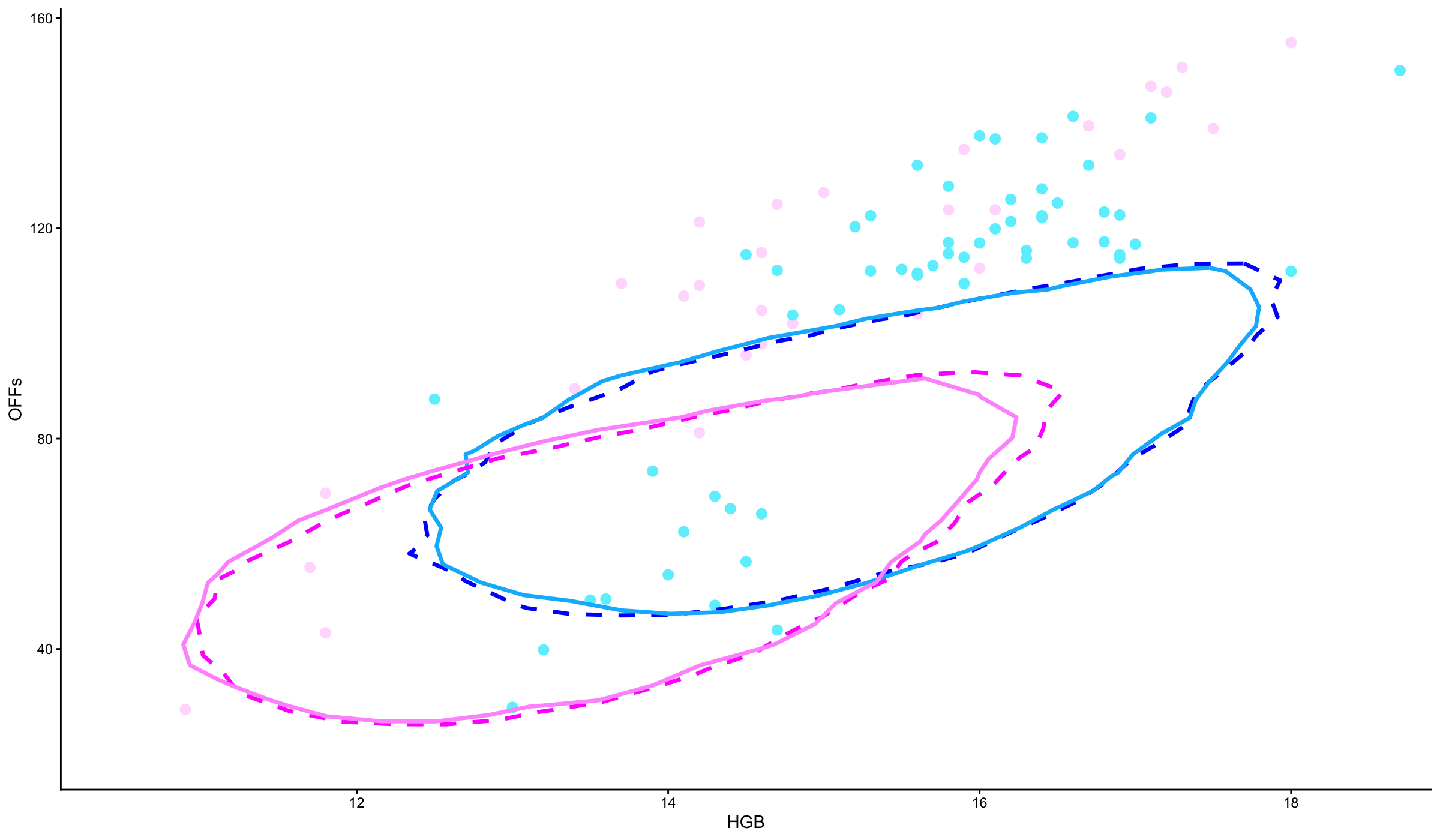}
    \caption{Estimated 95\% HPRs obtained with both the joint MCMC (dashed lines) and Bayesian IFM (solid lines) approaches. Points represent observations of athletes convicted of anti-doping rule violations. Blue and pink regions correspond to male and female athletes, respectively.}
    \label{fig: hpr doped}
\end{figure}

Figure~\ref{fig: hpr doped} shows the same 95\% HPRs together with observations from court rulings (\href{https://jurisprudence.tas-cas.org}{TAS/CAS}, \href{https://www.uci.org/judicial-bodies/4DeAGy6jHYGdec9R49wKVS}{UCI}) involving athletes convicted of anti-doping rule violations related to the hematological module of the ABP.
For female athletes, 89.7\% of the observations are classified as atypical, while for male athletes, joint MCMC identifies 82.5\% and Bayesian IFM 84.2\% as atypical. Overall, the two approaches classify 84.9\% and 86\% of the observations as atypical, respectively, as shown in Table~\ref{tab: tpr}.  \\

Notably, court rulings may report observations that are atypical with respect to the athlete's individual reference range or reflect suspicious variations in a biomarker over a specific time interval; not all reported observations are necessarily of this kind. We attempted to select only those identified as atypical based on the evidence reported in the court decisions; nevertheless, the sample may still contain observations that are not entirely relevant to the problem under investigation.
\begin{table}[!htbp]
    \centering
    \begin{tabular*}{\linewidth}{@{\extracolsep\fill}lccc@{\extracolsep\fill}}
    \toprule
        & \multicolumn{3}{c}{\bf TPR} \\ \cline{2-4} \addlinespace
        & Male & Female & {\bf Total} \\ 
    \midrule\addlinespace
        \textbf{joint MCMC} & 0.825 & 0.897 & 0.849 \\ \addlinespace
        \textbf{Bayesian IFM} & 0.842 & 0.897 & 0.860 \\
    \bottomrule
    \end{tabular*}
    \caption{Classification performance of the predictive regions obtained from the joint MCMC and Bayesian IFM approaches in terms of true positive rate (TPR).}
    \label{tab: tpr}
\end{table}

\section{Discussion and Conclusion}\label{sec: concl}

This work focuses on the Bayesian computational aspects of estimating multivariate copula-based models and was directly motivated by an application to doping detection with two primary biomarkers of the Athlete Biological Passport. Specifically, we introduced a fully joint MCMC alternative (Section~\ref{sec: full_MCMC}), which updates marginal and copula parameters within a single Markov chain, thereby allowing dependence information to flow in both directions. A side-by-side comparison was then carried out with the two-step Bayesian IFM strategy proposed in \cite{deliu2025multivariatecopulabasedbayesianframework} (Section~\ref{sec: Bayes_IFM}).

The comparative simulation study of Section~\ref{sec: simulations} showed that, under a correctly specified copula, both procedures recover the true parameters with comparably low bias and RMSE, with the joint MCMC exhibiting a modest advantage for the copula dependence parameter and Kendall's $\tau$. Under copula misspecification, by contrast, the two procedures displayed an asymmetric pattern for the two parameter blocks. For the marginal parameters, the Bayesian IFM performed markedly better than the joint MCMC, especially when the misspecified fitted copula was the Frank or the survival Clayton. This is a direct consequence of the modular, cut-feedback structure of the Bayesian IFM: by preventing information from flowing from the copula module back to the marginal module, it shields the marginal parameters from copula misspecification. The joint MCMC's marginal estimates, on the other hand, are updated jointly with the copula parameters at every iteration, so that a misspecified copula can contaminate the marginal inference through this shared estimation process. This result is consistent with the theoretical and empirical findings of \cite{smith2025cutting} for cut posteriors. For the dependence parameter ($\theta$ or $\tau$), both the Bayesian IFM and the joint MCMC deteriorated substantially once the fitted copula departed from the true one, and did so far more severely for the Bayesian IFM than for the joint MCMC. It is also worth noting that the two approaches showed some divergence in computational cost, with the joint MCMC showing remarkable advantages over the Bayesian IFM counterpart. 

The application to the NADO Italia dataset complemented the simulation findings in a real setting. Posterior estimates of the marginal and copula parameters obtained under the two procedures were broadly consistent, with the only notable discrepancy concerning the age effect on \texttt{OFFs}, for which the Bayesian IFM credible interval excluded zero while the joint MCMC interval did not. Both approaches produced comparable classification performance: a false positive rate of 4.0--4.5\% on the clean-athlete cohort, and a true positive rate of 85--86\% on a set of court-documented anti-doping rule violations, with no substantial advantage of one method over the other.

Taken together, these results point to a genuine trade-off rather than a uniformly preferable choice between the two procedures. The joint MCMC is markedly faster and provides better-calibrated inference for the dependence parameter, making it the natural default whenever computational cost is the binding constraint or when the copula family can be trusted to be reasonably well specified. The Bayesian IFM offers a practically important advantage in settings where the copula family is uncertain or difficult to validate. As emphasized throughout the anti-doping literature, false positives carry serious consequences for clean athletes, and since the marginal parameters directly determine the reference ranges against which an individual athlete's biomarker values are judged, their robustness to copula misspecification is not a secondary concern.

Several methodological extensions remain open for future work. First, the hierarchical marginal model of Section~\ref{sec: copula_mvt} (and related estimation) could be extended to accommodate individual random effects or temporal dependence within an athlete's longitudinal profile; while the assumption of time-independence may be reasonable for widely spaced measurements, within-individual serial dependence is likely to arise in practice. Second, although the present analysis focuses on the two primary hematological markers, the proposed framework extends naturally to a larger biomarker panel through vine copula constructions \citep{czado_analyzing_2019}, which represent high-dimensional dependence structures via bivariate building blocks; extending the Bayesian IFM to this setting would require estimating additional marginals and a cascade of conditional copulas, with a corresponding increase in computational complexity. In contrast, the joint MCMC would require only a proportional increase in the number of blocks updated at each iteration. Third, we note that MCMC is not the only computational route to joint Bayesian copula estimation. Variational inference has been proposed as a faster, approximate alternative for vine-copula-structured likelihoods \citep{kejzlar2023variational}, and approximate Bayesian computation has similarly been used to sidestep intractable copula likelihoods altogether \citep{grazian_approximate_2022}. We did not pursue either direction here, given the moderate dimensionality of our application, but both represent natural avenues for scaling the proposed framework to larger biomarker panels. Finally, the highest predictive regions used here to flag atypical profiles rely on asymptotic justifications for their coverage properties; an interesting alternative would be to construct such uncertainty regions using conformal inference methods (see \cite{vovk2005algorithmic} and \cite{deliu2026interplay} for a Bayesian treatment), which provide finite-sample coverage guaranties under comparatively mild assumptions and could complement or substitute the approach adopted here. 

Beyond its methodological contribution, this work has direct practical relevance for anti-doping practice. Complementing \cite{deliu2025multivariatecopulabasedbayesianframework}, who already showed the substantial advantage of the multivariate copula-based extension of ADAPTIVE, the present paper provides the technical machinery and a concrete comparison between two viable inferential routes, needed to actually implement this multivariate upgrade in practice. For international and national anti-doping organizations, this means the improved discriminating power of the multivariate model is no longer only a theoretical prospect: it comes with two fully specified, implementable estimation procedures, together with explicit guidance on the trade-offs between them, so that the choice can be tailored to the operational constraints and priorities of a given laboratory. In a setting where thousands of athlete profiles are screened routinely and the cost of a wrong decision is severe, having both a more accurate strategy and a viable route is what makes the transition to the multivariate ADAPTIVE model practically achievable.

\section*{Data Availability}
The data that support the findings of this study are owned by NADO Italia (National Anti-Doping Organisation Italy) and are not publicly available. They contain sensitive personal information on athletes and were provided exclusively for this research, under the explicit consent of the athletes involved. Restrictions therefore apply to the availability of these data, which were used under permission of NADO Italia for the current study. The observations of athletes convicted of anti-doping rule violations used in Section~\ref{sec: application} were retrieved from publicly available court rulings (TAS/CAS and UCI).

\section*{Acknowledgments}
The authors would like to thank NADO Italia, the owner of the dataset used in this study, which was collected as part of the project {\it Enhancing the methodological framework for doping detection: from univariate to multivariate testing analysis on Athlete Biological Passport profiles}. The project particularly involved Dr. Michaela Procaccini and Prof. Giuseppe D’Onofrio.

\section*{Financial Disclosure}
This project has been carried out with the support of the World Anti-Doping Agency and the Partnership for Clean Competition (WADA Grant 23D01BL). The funders had no role in the study design, data analysis, and interpretation of the results. 

\section*{Conflicts of Interest}

The authors declare no conflicts of interest.

\bibliography{main.bib}


\newpage

\appendix

\section{Model without covariates \label{appendix: no_cov}}%
In the model without covariates, the marginal parameters are estimated via maximum likelihood (MLE). 
We assume Gaussian distributions for both marginals, with $\mu \approx 14.14$ and $\sigma \approx 1.24$ for Hgb, and $\mu \approx 69.20$ and $\sigma \approx 17.15$ for OFFs. After transforming the observations to the unit interval, a Gumbel copula model is fitted to the data, yielding Kendall's $\tau = 0.55$ and a corresponding dependence parameter $\theta \approx 2.22$. The resulting data-generating process is defined as follows:
\begin{align}
    &Y_i^{\texttt{Hgb}} \sim \mathcal{N}(\mu = 14.14, \sigma = 1.24), \qquad i=1,\dots,n \nonumber \\
    &Y_i^{\texttt{OFFs}} \sim \mathcal{N}(\mu = 69.20,\, \sigma = 17.15) \\
    &(U_i^{\texttt{Hgb}}, U_i^{\texttt{OFFs}}) \sim C_\theta^\texttt{Gumbel} (\theta = 2.22). \nonumber
\end{align} 
Subsequently, the simulated observations are transformed back to their original scale using the inverse of the marginal CDF, $Y_i^{(\ell)} = \Phi^{-1}_{(\ell)}(U^{(\ell)}; \mu^{(\ell)}, \sigma^{(\ell)})$.
For each simulated dataset, a Gaussian marginal distribution is specified for both variables. The data are then standardized, and weakly informative priors are assigned to the marginal parameters
\begin{align}
    \mu^{(\ell)} &\sim \mathcal{N}(0, 2.5^2), \\
    \sigma^{(\ell)} &\sim \mathrm{half}-t_3(0, 2.5). \nonumber
\end{align}

Dependence is modeled through a Gumbel copula, with a uniform prior assigned to the dependence parameter $\theta$
\begin{align}
    (U_{i}^{\texttt{Hgb}}, U_{i}^{\texttt{OFFs}}) | \theta \overset{ind}{\sim} C_\theta^{\texttt{Gumbel}}, \quad \theta \sim \mathcal{U}(1,10).
\end{align}
The prior is restricted to positive dependence, given the intrinsic association between Hgb and OFFs. 

\begin{figure*}[!tbh]
    \centering
    \includegraphics[width=\linewidth]{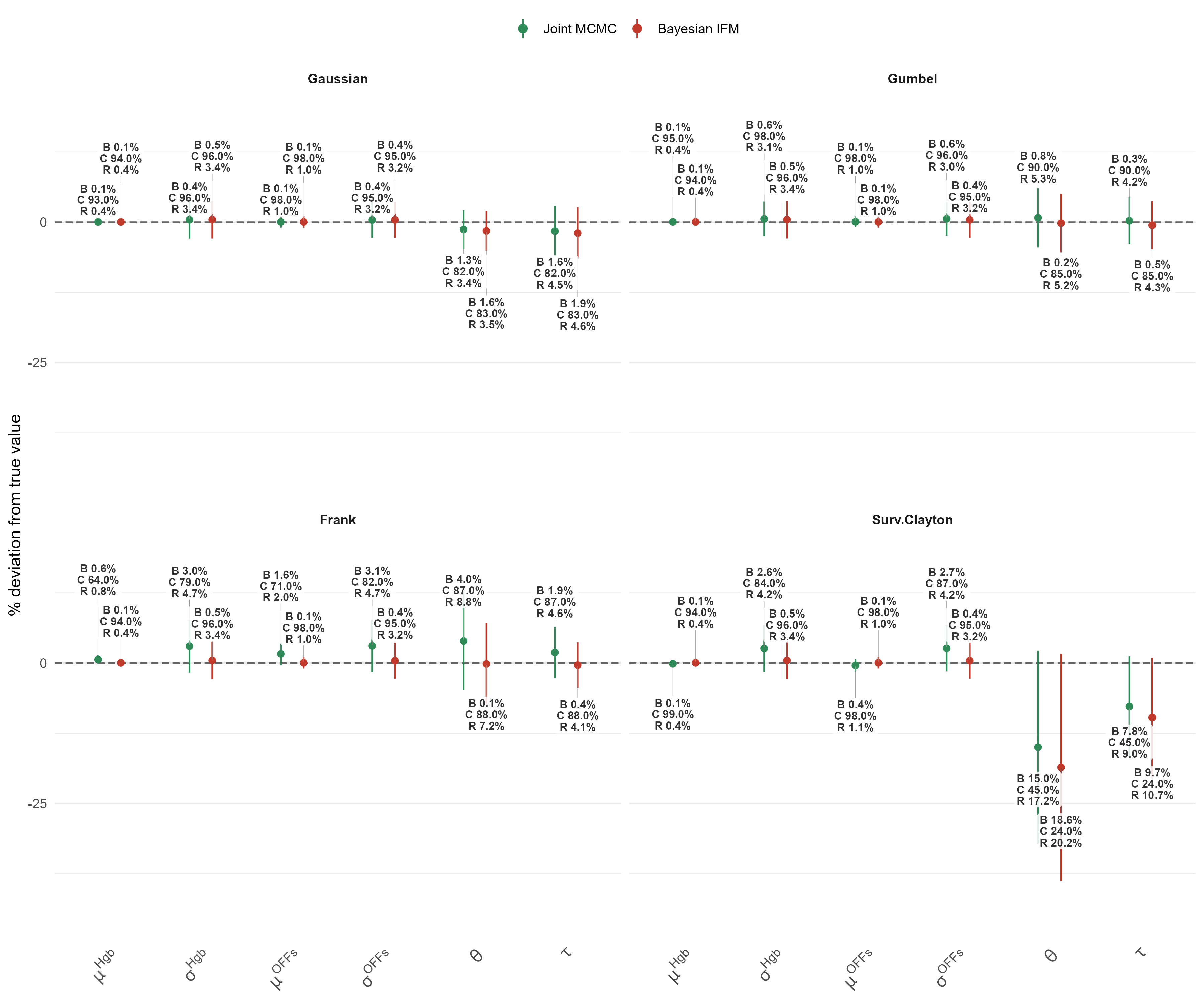}
    \caption{Points show the mean relative bias per parameter for the Joint MCMC (green) and Bayesian IFM (red) estimators, under the Gaussian (top-left), Gumbel (top-right), and Frank (bottom-left) and Surv. Clayton (bottom-right) copulas without covariates ($n = 500$, 100 simulations); The dashed line marks zero deviation and, labels give relative bias (B), coverage of the 95\% credible interval (C), and relative RMSE (R).}
    \label{fig:bias_all_copulas}
\end{figure*}

\subsection{Results}
The upper-right panel of Figure~\ref{fig:bias_all_copulas} shows the results obtained from the two approaches for the model without covariates.
For all parameters considered, the bias and RMSE values are relatively small, indicating 
a good ability of both approaches to recover the true parameter values.
The most notable differences arise for the copula dependence parameter and Kendall's $\tau$. Joint MCMC exhibits a positive bias and a coverage probability of 90\%, whereas Bayesian IFM shows a negative bias and a coverage probability of 85\%. These results suggest a slight overestimation of dependence for the former and an underestimation for the latter, together with undercoverage for both approaches, which is more pronounced for Bayesian IFM. In contrast, the coverage probabilities for the marginal parameters are close to or above the nominal level.

\subsection{Copula misspecification}
Three different misspecified copula families are considered: Gaussian, Frank and survival Clayton. Dependence is modeled through the parameter specific to each copula family, with the parameter space chosen to ensure coverage of Kendall's $\tau$ over the interval $(0, 0.9)$. 
Figure~\ref{fig:bias_all_copulas} reports the results under copula misspecification, including the corresponding bias, coverage probabilities, and RMSE. In all three cases, Bayesian IFM provides accurate estimates of the marginal parameters, with coverage probabilities close to the nominal level. Joint MCMC performs well under the Gaussian copula, whereas under Frank and survival Clayton it provides less accurate estimates, particularly for the scale parameters, with coverage probabilities below the nominal level.
For the copula parameters, Gaussian copula performs well, and the Frank copula shows a slightly larger bias only for Joint MCMC. For the survival Clayton copula, however, the bias increases substantially for both methods, with a marked decrease in coverage probability (45\% for Joint MCMC, 24\% for Bayesian IFM). 
Overall, these results indicate greater stability of Bayesian IFM in estimating the marginal parameters under copula misspecification, while inference on the dependence structure is more sensitive to the assumed copula family.

\section{Linear Model Fitting \label{appendix: fit}}%
Table \ref{tab: fit} reports the results of the linear models fitted to Hgb and OFFs. The estimated coefficients are used to define the marginal parameters of the data-generating process.

\begin{table}[!htbp] \centering 
\begin{tabular}{@{\extracolsep{5pt}}lcc} 
\\[-1.8ex]\hline 
\hline \\[-1.8ex] 
 & \multicolumn{2}{c}{\textit{Dependent variable:}} \\ 
\cline{2-3} 
\\[-1.8ex] & $\texttt{Hgb}$ & $\texttt{OFFs}$ \\ 
\hline \\[-1.8ex] 
  $\beta_0$ -- Constant & 14.178$^{***}$ & 51.503$^{***}$ \\ 
  & (0.232) & (3.180) \\ \addlinespace
  $\beta_1$ -- Gender & 1.536$^{***}$ & 22.292$^{***}$ \\ 
  & (0.127) & (1.740) \\ \addlinespace
  $\beta_2$ -- Age & $-$0.028$^{***}$ & 0.217$^{**}$ \\ 
  & (0.007) & (0.100) \\ [.6ex]
\hline \\[-1.8ex] 
Observations & 223 & 223 \\ 
R$^{2}$ & 0.421 & 0.435 \\ 
Adjusted R$^{2}$ & 0.416 & 0.430 \\ 
Residual Std. Error (df = 220) & 0.949 & 12.987 \\ 
F Statistic (df = 2; 220) & 80.010$^{***}$ & 84.603$^{***}$ \\ 
\hline 
\hline \\[-1.8ex] 
\textit{Note:}  & \multicolumn{2}{r}{$^{*}$p$<$0.1; $^{**}$p$<$0.05; $^{***}$p$<$0.01} \\ 
\end{tabular} 
\caption{Linear model fits for Hgb and OFFs, reporting $R^2$, adjusted $R^2$, residual standard errors, F-statistics, and significance levels for each covariate.} 
  \label{tab: fit} 
\end{table} 

\end{document}